\documentclass[twocolumn,10pt]{IEEEtran}
\usepackage{amsmath}
\usepackage{amsfonts}
\usepackage{epsfig}
\usepackage{amssymb}
\usepackage{cite}
\usepackage{color,soul}
\usepackage{subfigure}
\usepackage{multirow}
\usepackage{rotating}
\usepackage{graphicx}
\usepackage{tabularx}
\usepackage{array}
\usepackage{graphicx,dblfloatfix}
\usepackage{blindtext}
\usepackage{ragged2e}
\usepackage{xcolor,colortbl}
\definecolor{Gray}{gray}{0.85}
\usepackage{amsthm,amssymb,amsmath,bm}
\usepackage[subfigure]{tocloft}
\usepackage[font={small}]{caption}
\usepackage{algorithmic}
\usepackage[Algorithm]{algorithm}
\begin{document}
 \title{Covert Communication Using Null Space and 3D Beamforming}
\author{\IEEEauthorblockN{Moslem Forouzesh, Paeiz Azmi,
		Nader Mokari, 
		and Dennis Goeckel}}
  \vspace{-10cm}
 \maketitle

 \begin{abstract}
Covert communication is often limited in rate because it is difficult to hide the signal in the background noise.  Recent work has shown that jamming can significantly improve the rate at which covert communications can be conducted; however, the rate could be improved further if the jamming incident on the intended receiver can be mitigated.  Here, we consider a multiple-antenna jammer that employs beamforming to place the intended receiver in the null space of the jamming and a multi-antenna transmitter equipped with three-dimensional (3D) antennas that is able to beamform toward its intended recipient. To evaluate this design, we formulate an optimization problem and present an iterative algorithm to solve it.  Numerical results consider both the rate of covert communications with the proposed architecture and the gap between the result from our optimization and that obtained from exhaustive search.
 \end{abstract}
\vspace{-.7cm}
  \section{Introduction}
Security is a vital issue in wireless communications. In some cases, the protection of the message's content is not sufficient; rather, it is important to prevent the adversary from even understanding that the message exists.  This method which conceals communication between nodes is termed ``covert communication" and has been studied extensively in recent years, starting with a treatment of the order of the covert throughput on additive white Gaussian noise (AWGN) channels \cite{AWGN_isit, AWGN_ch}.   This was rapidly extended to discrete memoryless channels (DMCs) and the determination of the scaling constants \cite{Jaggi_ISIT, Bloch_IEEEIT, Wang_IEEEIT}.  In  \cite{Covert},
covert communication when users have uncertain channel state information (CSI) is studied.
Robust power allocation in covert communication under the assumption of imperfect channel distribution information (CDI) is investigated in \cite{Imperfect_CDI}.  Furthermore, the authors in \cite{FK} study communication with a covertness requirement in untrusted relaying networks. 

The results of \cite{AWGN_isit, AWGN_ch} indicate that covert communication is challenging on an AWGN channel: only $O(\sqrt{n})$ bits in $n$ channel uses can be transmitted covertly and reliably.  However, if the adversary is unaware of the background noise level, $O(n)$ bits can be transmitted \cite{baxley_journal}, and \cite{jammer} proposes an approach that employs a jammer only loosely coordinated with the transmitter that achieves such.  We also employ an external jammer here, which allows the transmitter to employ a significantly higher power while remaining covert.  But, if the jamming is sent with a single isotropic antenna as in \cite{jammer}, the jamming signal decreases the received signal-to-noise ratio (SNR) at the legitimate receiver.  In order to avoid this detriment, we propose to equip the jammer with multiple antennas and to employ null space beamforming.  Moreover, we exploit three-dimensional (3D) beamforming at the legitimate transmitter to focus the power of the message signal toward the legitimate receiver and potentially away from the adversary.  To our knowledge, this is the first investigation of how beamforming can improve the transmission rate of covert communications.
  
 The remainder of the paper is organized as follows: In Section \ref{Systam Model}, the system and signal model are presented. The detailed optimization problem and its solution are provided in Section \ref{Optimization Problem}. Section \ref{Numerical Results} presents numerical results. Finally, the conclusions of the work are presented in \ref{Conclusion}. 
\vspace{-.35cm}
 \section{System and Signal Model}\label{Systam Model}
\begin{figure}[t!]
 	\vspace{-1.3cm}
 	\hspace{-.1cm}	\includegraphics[width=3.3in,height=2in]{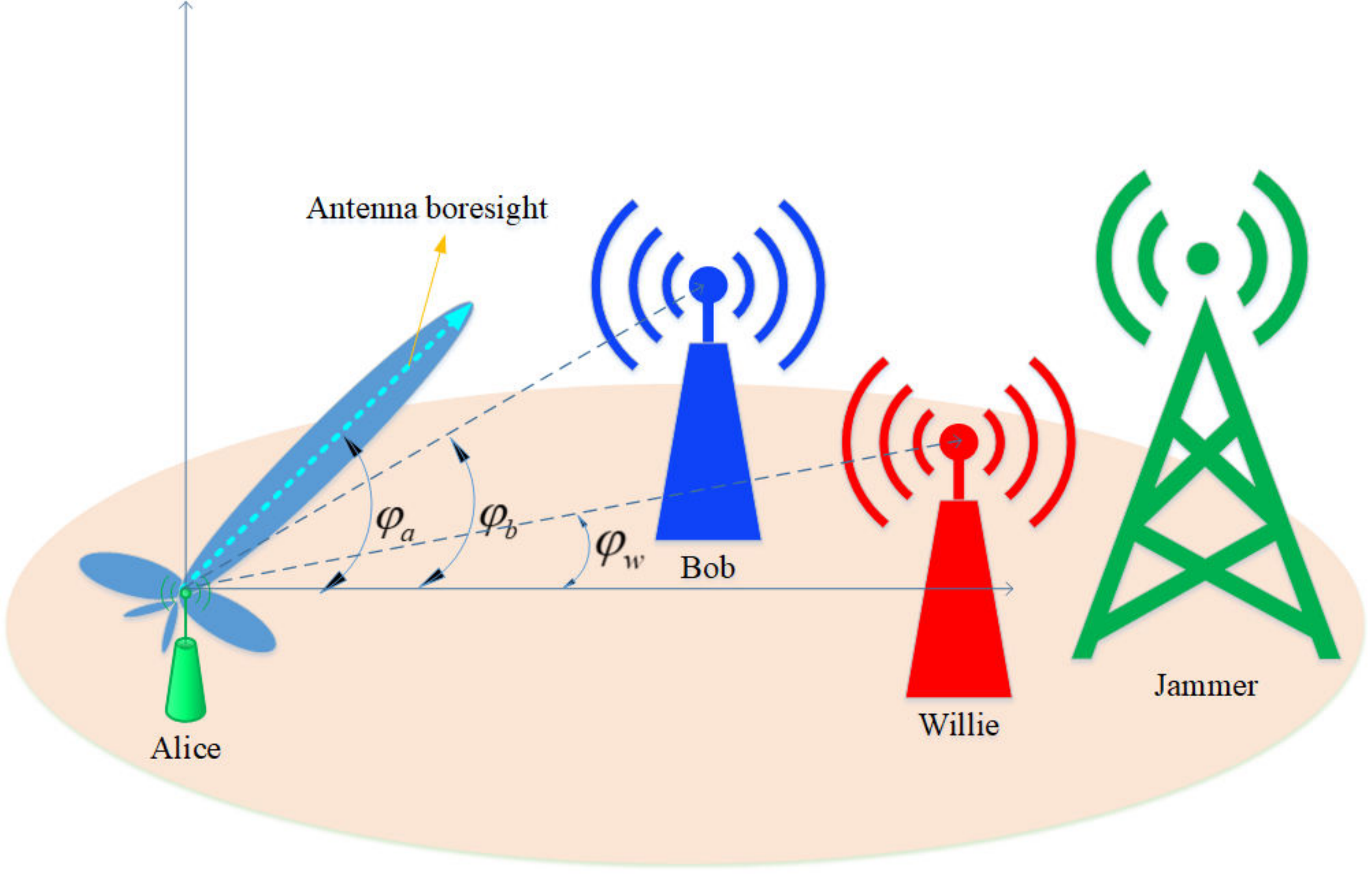}
 		\caption{System model.}
 		\label{System model}
 		\vspace{-.6cm}
 \end{figure}

 As seen in Fig \ref{System model}, the system model under investigation consists of a multiple-antenna legitimate transmitter (Alice) with $N_a$ antennas, a single-antenna legitimate receiver (Bob), a single-antenna warden (Willie), and a multiple-antenna jammer with $N_j$ antennas.  We assume the jammer is equipped with isotropic antennas and Alice is equipped with 3D antennas. 
  The aim of Alice is to transmit data to Bob covertly by employing three-dimensional (3D) beamforming, while the jammer injects a jamming signal to deceive Willie simultaneously.
  Because Willie is passive, we assume the CSI on channels from  system nodes to Willie is unknown; hence, the jammer is not able to focus power just on Willie receiver. Instead, to avoid the jamming signal inhibiting the legitimate receiver, the jammer employs a null-space beamforming technique. This communication is fulfilled through $T$ time slots; in each time slot, $n$ symbols are transmitted. 
 
The location of jammer,  Bob, and  Willie, are defined as  $(x_j, y_j)$, $(x_b, y_b)$, and $(x_w, y_w)$, respectively. Without loss generality, we assume Alice is located in $(0,0)$. The complex Gaussian channel vector from Alice to Bob, Alice to Willie,  jammer to Bob, and jammer to Willie  are defined by ${{\bf{h}}_{ab}} \sim {\cal C}{\cal N}\left( {{{\bf{0}}_{{N_a} \times 1}},{\boldsymbol{C}_{ab}}} \right)$, ${{\bf{h}}_{aw}} \sim {\cal C}{\cal N}\left( {{{\bf{0}}_{{N_a} \times 1}},{\boldsymbol{C}_{aw}}} \right)$, ${{\bf{h}}_{jb}} \sim {\cal C}{\cal N}\left( {{{\bf{0}}_{{N_j} \times 1}},{\boldsymbol{C}_{jb}}} \right)$, and ${{\bf{h}}_{jw}} \sim {\cal C}{\cal N}\left( {{{\bf{0}}_{{N_j} \times 1}},{\boldsymbol{C}_{jw}}} \right)$, where $\bf{0}$ is the zero matrix and $\boldsymbol{C}_{ij}$ is the positive definite channel covariance matrix between nodes $i$ and $j$. We assume that the channel vectors stay constant in each frame and vary independently across frames. 
 
In a 3D antenna, the antennas gain alters according to the transmit antenna pattern. The transmit antenna pattern is divided into a vertical and horizontal part \cite{ITU}.
 The horizontal antenna attenuation in dB scale is ${\Xi _H}\left( {{\theta _m}} \right) = \min \left\{ {12{{\left( {\frac{{{{\theta _m} - {\theta _0}}}}{{{\theta _{3dB}}}}} \right)}^2},{\Xi _m}} \right\}$ \cite{ITU},
   where $\Xi _m$ and $\theta _{3dB}$ are the maximum attenuation and the horizontal $3$dB beamwidth of Alice's antenna, respectively, $\theta  _0$ is the horizontal boresight angle of Alice, and $\theta  _m=\tan^{-1}(\frac{y_m}{x_m})$ (node $m$ is Willie or Bob). Without loss generality, we assume $\theta _0=0$.
Similar to the horizontal antenna attenuation the vertical  antenna attenuation can be written in  dB scale as 
$\Xi v\left( {{\varphi _m}} \right) = \min \left\{ {12{{\left( {\frac{{{\varphi _m} - {\varphi _a}}}{{{\varphi _{3dB}}}}} \right)}^2},{\Xi _m}} \right\}$ 
where $\varphi _a$ and $\varphi _{3dB}$ are the vertical boresight angle 
and the vertical  $3$dB beamwidth of Alice's antenna. Furthermore, ${\varphi _m} = {\tan ^{ - 1}}\frac{{\Delta {z_m}}}{{\sqrt {{x_m} + {y_m}} }}$, where $\Delta {z_m}$ is the height difference between Alice and node $m$. Finally, the antenna gain in dB scale is formulated as \cite{ITU}:
\begin{align}
\hspace{-.3cm}\Omega \left( {{\theta _m},{\varphi _m}} \right)=\Omega_{\max}-\min\left\{{\Xi v\left( {{\varphi _m}} \right)+{\Xi _H}\left( {{\theta _m}} \right),\Xi _m}\right\}.
\end{align}
where $\Omega_{\max}$ is the maximum antenna gain at boresight.
Given $\Xi v\left( {{\varphi _m}} \right)+{\Xi _H}\left( {{\theta _m}} \right) \le \Xi _m $, the antenna gain in
linear scale can be reformulated as
$
\omega \left( {{\theta _m},{\varphi _m}} \right) = {10^{\frac{{{\Omega _{\max }}}}{{10}}}} \times {10^{ - 1.2\left[ {{{\left( {\frac{{{\theta _m} - {\theta _0}}}{{{\theta _{3dB}}}}} \right)}^2} + {{\left( {\frac{{{\varphi _m} - {\varphi _a}}}{{{\varphi _{3dB}}}}} \right)}^2}} \right]}}.
$

The jamming signal vector and Alice's message signal vector in each  time slot can be expressed as ${\bf{s}}_j = \left[ {s_j^{1 },s_j^{2 }, \ldots ,s_j^{n }} \right]$,  ${\bf{s}}_a = \left[ {s_a^{1 },s_a^{2 }, \ldots ,s_a^{n }} \right]$. 
The transmitted signal by  Alice and the jammer are 
$
\boldsymbol{X}_a^\ell={\bf{w}}_a s_a^\ell,
$
and
$
\boldsymbol{X}_j^\ell={\bf{w}}_js_j^\ell,
$
respectively, where ${\bf{w}}_a\in \mathbb{C}^{N_a\times 1}$ and ${\bf{w}}_j\in \mathbb{C}^{N_j\times 1}$ are beamforming vectors for transmission of the message and the jamming signal, respectively, to be designed in Section \ref{Optimization Problem}. Without loss of generality, we assume
 $\mathbb{E}\{ \left\| {{s_a^\ell}} \right\|^2\}  = 1$ and  $\mathbb{E}\{ \left\| {{s_j^\ell}} \right\|^2\}  = 1$, where $\mathbb{E}$ is the expectation operator. Per above, CSI for channels from system nodes to Willie is unknown at Alice, but we will  assume the CDI (e.g., fading environment) is available. We assume that the CSI for channels from Alice and the jammer to Bob is available.

 The  $\ell^{th}$ received signal at $m$ in each time
 slot is given by  
 \begin{align}
 &{{y}_m^\ell} = \\&\hspace{-.15cm}\left\{ {\begin{array}{*{20}{l}}
 	\hspace{-.25cm}{d_{jm}^{-\alpha/2}{\bf{h}}^H _{jm}{{\bf{w}}_j}s_j^{\ell} + {n_m^\ell},}&\hspace{-.3cm}{{\Psi _0},}\\
 	\hspace{-.25cm}{d_{am}^{-\alpha/2}{\sqrt{\omega\left(\theta_b,\varphi_b\right)p_a}}{\bf{h}}^H_{am}{{\bf{w}}_a} s_a^{\ell} + d_{jm}^{-\alpha/2} {\bf{h}}^H_{jm}{{\bf{w}}_j}s_j^{\ell} +n_m^\ell,}&\hspace{-.3cm}{{\Psi _1},}
 	\end{array}} \right.\nonumber
 \end{align}
 where $\alpha$, $d_{am}$, $d_{jm}$, and $(.)^H$ are the path loss exponent, distance between Alice and node $m$, distance between the jammer and node $m$, and hermitian operator respectively. Also, ${n_m^\ell} \sim {\cal C}{\cal N}\left( {{0},{\sigma_m^2}} \right)$ is the received additive white Gaussian noise at node $m$. Since Alice has perfect knowledge of the CSI on channels to Bob, she is able to employ the well known
 Maximum Ratio Transmission technique (MRT), i.e., ${\bf{w}}_a={\bf{h}}_{ab}/{ \left\| {{\bf{h}}_{ab}} \right\|}$.
\subsection{Covert Communication Requirement}
$\Psi _0$ indicates Alice does not transmit data and $\Psi _1$ indicates Alice  transmits data.  
Establishing the optimal receiver at Willie can be challenging for complicated environments. For simple environments with or without jamming, Willie's optimal receiver is often a radiometer; that is, Willie compares the total received power to a threshold 
[1,2,10]. Although establishing the optimality of a radiometer is difficult here, it is the receiver that would \textbf{most} likely be employed in practice and is thus often assumed (e.g., \cite{baxley_journal}). Hence, we will assume that Willie employs a radiometer and thus employs the test $\frac{{r_w}}{n} \mathop \gtrless\limits_{\Psi_0}^{\Psi_1}\vartheta$, where $r_w={\sum\limits_{\ell   = 1}^n {\left| {y_w^\ell  } \right|} ^2}$, and $\vartheta$ is the decision threshold.
When Alice transmits data but Willie decides $\Psi_0$, a missed detection (MD) with probability $\mathbb{P}_{MD}$ has occurred, when Alice does not transmit but Willie decides $\Psi_1$, a false alarm (FA) with probability $\mathbb{P}_{FA}$ has occurred. Finally, we say this communication is covert when
``$\text{for any}\, \varepsilon \ge 0, \,\,\, \mathbb{P}_{MD}+\mathbb{P}_{FA}\ge 1-\varepsilon, \,\,\, \text{as}\,\,\, n \to \infty,
$'' is satisfied \cite{AWGN_isit,AWGN_ch}
In this paper, it is assumed that Alice  employs Gaussian codebooks and the jammer uses Gaussian jamming; \cite{jammer} hence,
$
y_w^\ell  \mid h_{aw},h_{jw}  \sim   {\cal C}{\cal N}\left( {0,\sigma _w^2 + \gamma } \right)
$ and  ${\sum\limits_{\ell   = 1}^n {\left| {y_w^\ell  } \right|} ^2}\mid h_{aw},h_{jw}  \sim \left( {\sigma _w^2 + \gamma } \right) \chi _{2n}^2$, where $\chi _{2n}^2$ is  chi-squared random variable with $2n$ degrees of freedom
and
\begin{align}
 \gamma = \left\{ {\begin{array}{*{20}{l}}
	\hspace{-.28cm}{d_{jw}^{-\alpha}{\left| {{{\bf{w}}^H_j} {\bf{h}}_{jw}} \right|^2}},&\hspace{-.35cm}{{\Psi_0}},\\
	\hspace{-.28cm}{d_{aw}^{-\alpha}{{\omega\left(\theta_w,\varphi_w\right)p_a}}{\left| {{\frac{{\bf{h}}^H_{ab}{\bf{h}}_{aw}}{ \left\| {{\bf{h}}_{ab}} \right\|}}}‌ \right|^2} +d_{jw}^{-\alpha}{\left| {{{\bf{w}}^H_j} {\bf{h}}_{jw}} \right|^2}},&\hspace{-.35cm}{{\Psi_1}}.
	\end{array}} \right.
\end{align}
 The FA and MD  probabilities are given by \cite{Covert}:
\begin{equation}\label{PFA}
{\mathbb{P}_{FA}} = \mathbb{P}\left( {\frac{{{Y_w}}}{n} > \vartheta \left| {{\Psi_0}} \right.} \right) = \mathbb{P}\left( {\left( {\sigma _w^2 + \gamma } \right)\frac{{\chi _{2n}^2}}{n} > \vartheta \left| {{\Psi_0}} \right.} \right),
\end{equation}
\begin{equation}\label{PMD}
{\mathbb{P}_{MD}} = \mathbb{P}\left( {\frac{{{Y_w}}}{n} < \vartheta \left| {{\Psi_1}} \right.} \right) = \mathbb{P}\left( {\left( {\sigma _w^2 + \gamma } \right)\frac{{\chi _{2n}^2}}{n} < \vartheta \left| {{\Psi_1}} \right.} \right),
\end{equation}
We take an outage approach here: we let $n \to \infty$ and consider probability that channel conditions are such that covert communication is accomplished \cite{Covert}.
 According to the Strong Law of Large Numbers (SLLN),
 $\frac{{\chi _{2n}^2}}{n}$  converges to 1, and based on 
 Lebesgue's Dominated Convergence Theorem, we are able to replace  $\frac{{\chi _{2n}^2}}{n}$  with 1 when $n \to \infty $. Hence, we can write  the FA and MD  probabilities as 
$
{\mathbb{P}_{FA}} = \mathbb{P}\left( { {\sigma _w^2 + \gamma } > \vartheta \left| {{\Psi_0}} \right.} \right),\,
 {\mathbb{P}_{MD}}= \mathbb{P}\left( { {\sigma _w^2 + \gamma } < \vartheta \left| {{\Psi_1}} \right.} \right)
$. The probability density function of $\gamma$ is
\begin{equation}\label{f_g}
{f_\Gamma }\left( \gamma  \right) = \left\{ {\begin{array}{*{20}{l}}
	{\frac{{{e^{ - \frac{\gamma }{{d_{jw}^{ - \alpha }{\bf{w}}_j^H{{\bf{C}}_{jw}}{{\bf{w}}_j}}}}}}}{{d_{jw}^{ - \alpha }{\bf{w}}_j^H{{\bf{C}}_{jw}}{{\bf{w}}_j}}},}&{{\Psi _0},}\\
	{\frac{{{e^{ - \frac{\gamma }{{d_{aw}^{ - \alpha }\omega \left( {{\theta _w},{\varphi _w}} \right){p_a}{\bf{w}}_a^H{{\bf{C}}_{aw}}{{\bf{w}}_a}}}}} - {e^{ - \frac{\gamma }{{d_{jw}^{ - \alpha }{\bf{w}}_j^H{{\bf{C}}_{jw}}{{\bf{w}}_j}}}}}}}{{d_{aw}^{ - \alpha }\omega \left( {{\theta _w},{\varphi _w}} \right){p_a}{\bf{w}}_a^H{{\bf{C}}_{aw}}{{\bf{w}}_a} - d_{jw}^{ - \alpha }{\bf{w}}_j^H{{\bf{C}}_{jw}}{{\bf{w}}_j}}},}&{{\Psi _1}.}
	\end{array}} \right.
\end{equation}
By employing  \eqref{f_g}, ${\mathbb{P}_{FA}}$ and ${\mathbb{P}_{MD}}$ can be calculated as follows:
\begin{equation}
\hspace{-2.5cm}{\mathbb{P}_{FA}} = \left\{ {\begin{array}{*{20}{l}}
	{{e^{ - \frac{{\left( {\vartheta  - \sigma _w^2} \right)}}{{d_{jw}^{ - \alpha }{\bf{w}}_j^H{{\bf{C}}_{jw}}{{\bf{w}}_j}}}}},}&{\vartheta  - \sigma _w^2 > 0,}\\
	{1,}&{\vartheta  - \sigma _w^2 \le 0.}
	\end{array}} \right.
\end{equation}
\begin{align}
&\mathbb{P}_{MD}=\\&\hspace{-.3cm}\left\{ {\begin{array}{*{20}{l}}
	\begin{array}{l}\hspace{-.4cm}
	1 + \frac{1}{{d_{aw}^{ - \alpha }\omega \left( {{\theta _w},{\varphi _w}} \right){p_a}{\bf{w}}_a^H{{\bf{C}}_{aw}}{{\bf{w}}_a} - d_{jw}^{ - \alpha }{\bf{w}}_j^H{{\bf{C}}_{jw}}{{\bf{w}}_j}}}\times\\\hspace{-.4cm}
	\left[ { + d_{jw}^{ - \alpha }{\bf{w}}_j^H{{\bf{C}}_{jw}}{{\bf{w}}_j}{e^{ - \frac{{\left( {\vartheta  - \sigma _w^2} \right)}}{{d_{jw}^{ - \alpha }{\bf{w}}_j^H{{\bf{C}}_{jw}}{{\bf{w}}_j}}}}} - d_{aw}^{ - \alpha }\omega \left( {{\theta _w},{\varphi _w}} \right)} \right.\\
	\left. { \hspace{-.4cm}\times {p_a}{\bf{w}}_a^H{{\bf{C}}_{aw}}{{\bf{w}}_a}{e^{ - \frac{{\left( {\vartheta  - \sigma _w^2} \right)}}{{d_{aw}^{ - \alpha }\omega \left( {{\theta _w},{\varphi _w}} \right){p_a}{\bf{w}}_a^H{{\bf{C}}_{aw}}{{\bf{w}}_a}}}}}} \right],{\hspace{.6cm}\vartheta  - \sigma _w^2 > 0,}
	\end{array}\\
	0,\hspace{6.5cm}{\vartheta  - \sigma _w^2 < 0}
	\end{array}} \right.\nonumber
\end{align}

\vspace{-.5cm}
\section{Optimization Problem} \label{Optimization Problem}

 Our aim is to maximize the covert rate at  Bob by setting
  the angle of Alice's  3D antennas and the null space beamforming vector at the jammer, subject to transmit power limitations and  covert communication requirements at  Bob. Hence, we consider the following optimization  problem:
 \vspace{-.24cm}
\begin{subequations}\label{optimization}
	\begin{align}
	&\max_{{\bf{w}}_j, p_a, \varphi_a}\; \mathbb{P}_t\log \left( {1 + \frac{{d_{ab}^{ - \alpha }{p_a}\omega \left( {{\theta _b},{\varphi _b}} \right)\mathrm{Tr}\left( {{{\bf{H}}_{ab}}} \right)}}{{d_{jb}^{ - \alpha }{{\left| {{\bf{w}}_j^H{{\bf{h}}_{jb}}} \right|}^2} + \sigma _b^2}}} \right),\\& \label{C_1}
	\hspace{.1cm}\text{s.t.}: p_a \le P_a^{\max },  \\&\label{C_2}
	\hspace{.78cm} \mathrm{Tr}\left( {{{{\bf{w}}}_j}{\bf{w}}_j^H} \right) \le P_j^{\max },
	\hspace{.008cm}
	\\&\label{C_3}
	\hspace{.78cm}
	\hspace{.008cm}{\mathrm{Tr}}\left( {{{\bf{w}}_j}{{\bf{h}}_{jb}^H}{\bf{h}}_{jb}{\bf{w}}_j^H} \right)=0,
	\\&\label{C_4}
	\hspace{.78cm}
	\hspace{.008cm}\min_{\vartheta}\left(\;{\mathbb{P}_{FA}} + {\mathbb{P}_{MD}}\right)\ge 1-\varepsilon,	
	\end{align}
\end{subequations}
where ${\bf{H}}_{ab}={\bf{h}}_{ab}{\bf{h}}_{ab}^H$, $\mathrm{Tr}(.)$ is the trace operator, and $\mathbb{P}_t$ is the probability of data transmission in each time slot.
For solving the optimization problem \eqref{optimization}, first we solve $\mathop {\min }\limits_\vartheta \left(\;{\mathbb{P}_{FA}} + {\mathbb{P}_{MD}}\right)$ to obtain the optimal $\vartheta $, in Willie's points of view; then, we try to solve \eqref{optimization} based on $\vartheta_{op}$.
\subsection{Optimal decision threshold}
Since Willie's aim is to minimize  $\mathbb{P}_{FA} + \mathbb{P}_{MD}$, he does not choose a value of  $\vartheta$ such that $\vartheta < \sigma_w^2$, because in this case $\mathbb{P}_{FA} + \mathbb{P}_{MD}=1$. Hence, we can consider the expression for the case $\vartheta > \sigma_w^2$. Accordingly, in order to obtain $\vartheta_{opt}$, we set $\frac{{\partial \left(\mathbb{P}_{FA} + \mathbb{P}_{MD}\right)}}{{\partial \vartheta}}=0$, which leads to the following result
	\begin{align}
{\vartheta _{op}} =& \frac{{\left( {d_{jw}^{ - \alpha }{\bf{w}}_j^H{{\bf{C}}_{jw}}{{\bf{w}}_j}} \right)\left( {d_{aw}^{ - \alpha }\omega \left( {{\theta _w},{\varphi _w}} \right){p_a}{\bf{w}}_a^H{{\bf{C}}_{aw}}{{\bf{w}}_a}} \right)}}{{d_{aw}^{ - \alpha }\omega \left( {{\theta _w},{\varphi _w}} \right){p_a}{\bf{w}}_a^H{{\bf{C}}_{aw}}{{\bf{w}}_a} - d_{jw}^{ - \alpha }{\bf{w}}_j^H{{\bf{C}}_{jw}}{{\bf{w}}_j}}}\times \nonumber\\&\ln \left( {\frac{{d_{aw}^{ - \alpha }\omega \left( {{\theta _w},{\varphi _w}} \right){p_a}{\bf{w}}_a^H{{\bf{C}}_{aw}}{{\bf{w}}_a}}}{{d_{jw}^{ - \alpha }{\bf{w}}_j^H{{\bf{C}}_{jw}}{{\bf{w}}_j}}}} \right)
\end{align}
\subsection{Solution of the Optimization  Problem}
The optimization problem \eqref{optimization} is non-convex; hence, we cannot use existing convex optimization methods. 
Instead, we employ the well-known iterative algorithm  called Alternative Search Method (ASM). In this algorithm, we convert this problem to two subproblems. In one subproblem, we find $p_a, {\bf{w}}_j$, and in the other, we find $\varphi_a$.
 First, we consider finding $p_a$ and $ {\bf{w}}_j$ given $\varphi_a$. The optimization problem is non-convex with respect to $p_a$ and $ {\bf{w}}_j$. We assume the covariance matrices are diagonal; hence,
\begin{align}\label{equvalent}
&\hspace{-.2cm}{\bf{w}}_j^H{{\bf{C}}_{jw}}{{\bf{w}}_j} = \mathrm{Tr}\left( {{\bf{w}}_j^H{{\bf{C}}_{jw}}{{\bf{w}}_j}} \right) = \mathrm{Tr}\left( {{{\bf{w}}_j}{\bf{w}}_j^H{{\bf{C}}_{jw}}} \right) = \sigma _{{h_{jw}}}^2\hspace{-.2cm}\mathrm{Tr}\nonumber\\&\hspace{-.2cm}\left( {{{\bf{w}}_j}{\bf{w}}_j^H{{\rm{I}}_{N_j \times N_j}}} \right) = \sigma _{{h_{jw}}}^2\mathrm{Tr}\left( {{{\bf{w}}_j}{\bf{w}}_j^H} \right) = \sigma _{{h_{jw}}}^2\mathrm{Tr}\left( {{{\bf{W}}_j}} \right),
\end{align}
 where ${\bf{W}}_j={\bf{w}}_j {{\bf{w}}_j^H}$. By defining  ${\bf{W}}_a={\bf{w}}_a {{\bf{w}}_a^H}$, we have \eqref{equvalent} for  ${\bf{w}}_a^H{{\bf{C}}_{aw}}{{\bf{w}}_a}$. Moreover,  by definition of an auxiliary variable $t$ and some mathematical manipulations, the optimization problem \eqref{optimization} can be rewritten as:
 \begin{subequations}\label{power}
 	\begin{align}
 	&\max_{{\bf{W}}_j, p_a, t}\; \mathbb{P}_t \log \left( {1 + \frac{{d_{ab}^{ - \alpha }{p_a}\omega \left( {{\theta _b},{\varphi _b}} \right)\mathrm{Tr}\left( {{{\bf{H}}_{ab}}} \right)}}{ \sigma _b^2}} \right),\\& \label{}
 	\hspace{-.5cm}\text{s.t.}:\hspace{.18cm}  p_a \le P_a^{\max },  \\&
 	\hspace{.45cm}{\mathrm{Tr}}\left( {{\bf{W}}_j} \right) \le P_j^{\max },
 	\hspace{.008cm}
 	\\&\label{}
 	\hspace{.45cm}
 	\hspace{.008cm}{\mathrm{Tr}}\left( {{{\bf{W}}_j}{{\bf{H}}_{jb}}} \right)=0,
 	 \\&
 	\label{} 
 	\hspace{.45cm}
 	\hspace{.008cm}
 	d_{aw}^{ - \alpha }\sigma _{{h_{aw}}}^2\omega \left( {{\theta _w},{\varphi _w}} \right){p_a} - d_{jw}^{ - \alpha }\sigma _{{h_{jw}}}^2{\mathrm{Tr}}\left( {{{\bf{W}}_j}} \right) \le t,
 	\end{align}
 	\begin{align}&\label{} 
 	\hspace{.45cm}
 	\hspace{.008cm}\ln \left( {\frac{{d_{jw}^{ - \alpha }\sigma _{{h_{jw}}}^2{\mathrm{Tr}}\left( {{{\bf{W}}_j}} \right)}}{{d_{aw}^{ - \alpha }\sigma _{{h_{aw}}}^2\omega \left( {{\theta _w},{\varphi _w}} \right){p_a}}}} \right)d_{jw}^{ - \alpha }\sigma _{{h_{jw}}}^2{\mathrm{Tr}}\left( {{{\bf{W}}_j}} \right) \\&\hspace{.55cm} - t\ln \left( \varepsilon  \right) \le 0,	\nonumber
 	\end{align}
 	 \end{subequations}
 where ${\bf{H}}_{jb}={\bf{h}}_{jb}{\bf{h}}_{jb}^H$. 
Finally, the optimization problem \eqref{power} is convex; hence, we  can employ available software such as the CVX solver \cite{CVX}. Note that when we have ${\bf{W}}_j$, we can obtain ${\bf{w}}_j$ by Gaussian randomization procedure \cite{SDR}.

Next, we find $\varphi_a$ given $p_a$ and $ {\bf{w}}_j$. In order to obtain  $\varphi_a$, we approximate ${\omega \left( {{\theta _w},{\varphi _w}} \right)}$ with respect to $\varphi_a$ as follows:
\begin{align}
&\omega \left( {{\theta _m},{\varphi _m},{\varphi _a}} \right) \simeq \tilde \omega \left( {{\theta _m},{\varphi _m},{\varphi _a}} \right) = \omega \left( {{\theta _m},{\varphi _m},{\varphi _a}\left( {k - 1} \right)} \right)\nonumber\\& + \nabla \omega \left( {{\theta _m},{\varphi _m},{\varphi _a}\left( {k - 1} \right)} \right)\left( {{\varphi _a}\left( k \right) - {\varphi _a}\left( {k - 1} \right)} \right)
\end{align}
where $\nabla$ is the gradient operator. Hence, we have 
\begin{align}
 &\nabla \omega \left( {{\theta _m},{\varphi _m},{\varphi _a}\left( {k - 1} \right)} \right) = {10^{\frac{{{\Omega _{\max }}}}{{10}}}} \times \\& \frac{{\left( {2.4 \times {{10}^{ - 1.2\left[ {{{\left( {\frac{{{\varphi _w} - {\varphi _a}(k - 1)}}{{{\varphi _{3dB}}}}} \right)}^2}} \right]}}\left( {{\varphi _w} - {\varphi _a}(k - 1)} \right)\ln \left( {10} \right)} \right)}}{{{{10}^{ + 1.2{{\left( {\frac{{{\theta _w} - {\theta _0}}}{{{\theta _{3dB}}}}} \right)}^2}}}\varphi _{3dB}^2}}.\nonumber
 \end{align}
  With this approximation, the optimization problem \eqref{power} is converted  to a convex problem with respect to $\varphi_a$, which is formulated as follows:
  \begin{subequations}\label{phi}
 	\begin{align}
 	&\max_{\varphi_a}\; \mathbb{P}_t \log \left( {1 + \frac{{d_{ab}^{ - \alpha }{p_a}\tilde \omega \left( {{\theta _m},{\varphi _m},{\varphi _a}} \right)\mathrm{Tr}\left( {{{\bf{H}}_{ab}}} \right)}}{ \sigma _b^2}} \right),\\& \label{}
 \text{s.t.}:
 	d_{aw}^{ - \alpha }\sigma _{{h_{aw}}}^2\tilde \omega \left( {{\theta _m},{\varphi _m},{\varphi _a}} \right){p_a} - d_{jw}^{ - \alpha }\sigma _{{h_{jw}}}^2{\mathrm{Tr}}\left( {{{\bf{W}}_j}} \right) \le t,
 	\\&\label{} 
 	\hspace{.2cm}\ln \left( {\frac{{d_{jw}^{ - \alpha }\sigma _{{h_{jw}}}^2{\mathrm{Tr}}\left( {{{\bf{W}}_j}} \right)}}{{d_{aw}^{ - \alpha }\sigma _{{h_{aw}}}^2\tilde \omega \left({{\theta _m},{\varphi _m},{\varphi _a}} \right){p_a}}}} \right)d_{jw}^{ - \alpha }\sigma _{{h_{jw}}}^2{\mathrm{Tr}}\left( {{{\bf{W}}_j}} \right) \nonumber\\&\hspace{.55cm} - t\ln \left( \varepsilon  \right) \le 0.
 	\end{align}
 \end{subequations}
  Since we have converted the main optimization problem to two subproblems, we propose an iterative algorithm as shown in Algorithm I. 
  \begin{algorithm}[t] 
	\caption{PROPOSED ITERATIVE ALGORITHM} \label{Alg_M1}
	\begin{algorithmic}[1]
		\STATE  \nonumber
		Initialization: Set $k =0$ ($k$ is the iteration number)
		and initialize to  $\varphi_a(0)$,
		\STATE 
		Set $\varphi_a=\varphi_a( k)$,
		\STATE  		
		Solve the convex optimization problem \eqref{power} given $\varphi_a$ to obtain $p_a$ and $ {\bf{w}}_j$; then, set  the outcomes  to $p_a(k+1)$ and ${\bf{w}}_j(k+1)$, respectively.
		\STATE 
		Solve \eqref{phi} given $p_a(k+1)$ and ${\bf{w}}_j(k+1)$ and set  the outcome  to $\varphi_a(k+1)$.
		\STATE 
		If $\left| {R (k + 1) - R (k)} \right| \le \tau $, ($\tau$ is the stopping threshold and $R$ is the objective function),\\
		stop; \\
		Else:
		Set $k = k + 1$ and go back to step 2.
	\end{algorithmic}
\end{algorithm}
\vspace{-.2cm}
\section{Numerical Results} \label{Numerical Results}
In this section, we evaluate the proposed scheme. The simulation settings are: $\varepsilon=0.1$, $\Omega_{\max}=17$ dB, $\varphi _{3dB}= 10^\circ $, $\theta _{3dB}= 70^\circ $, $\alpha=4$, $\sigma_b^2=\sigma_w^2=-30$ dBW, $N_j=8$, $\mathbb{P}_t=0.5 $, $\Delta {z_m}=7.5$ m (meter), and the location of nodes are ${L_a} = (0,0),{L_j} = (0, - 10),{L_b} = (10,0),{L_w} = (x_w, 0)$.

\begin{figure*}[t]
	\begin{center}$
		\begin{array}{cc}
		\hspace{-0.45cm}	\subfigure[h][Covert rate versus $N_a$, $P_j^{\max }=P_a^{\max }=0.5$ Watts and $x_w=8$ m.]{
			\includegraphics[width=2.4 in,height=1.8in]{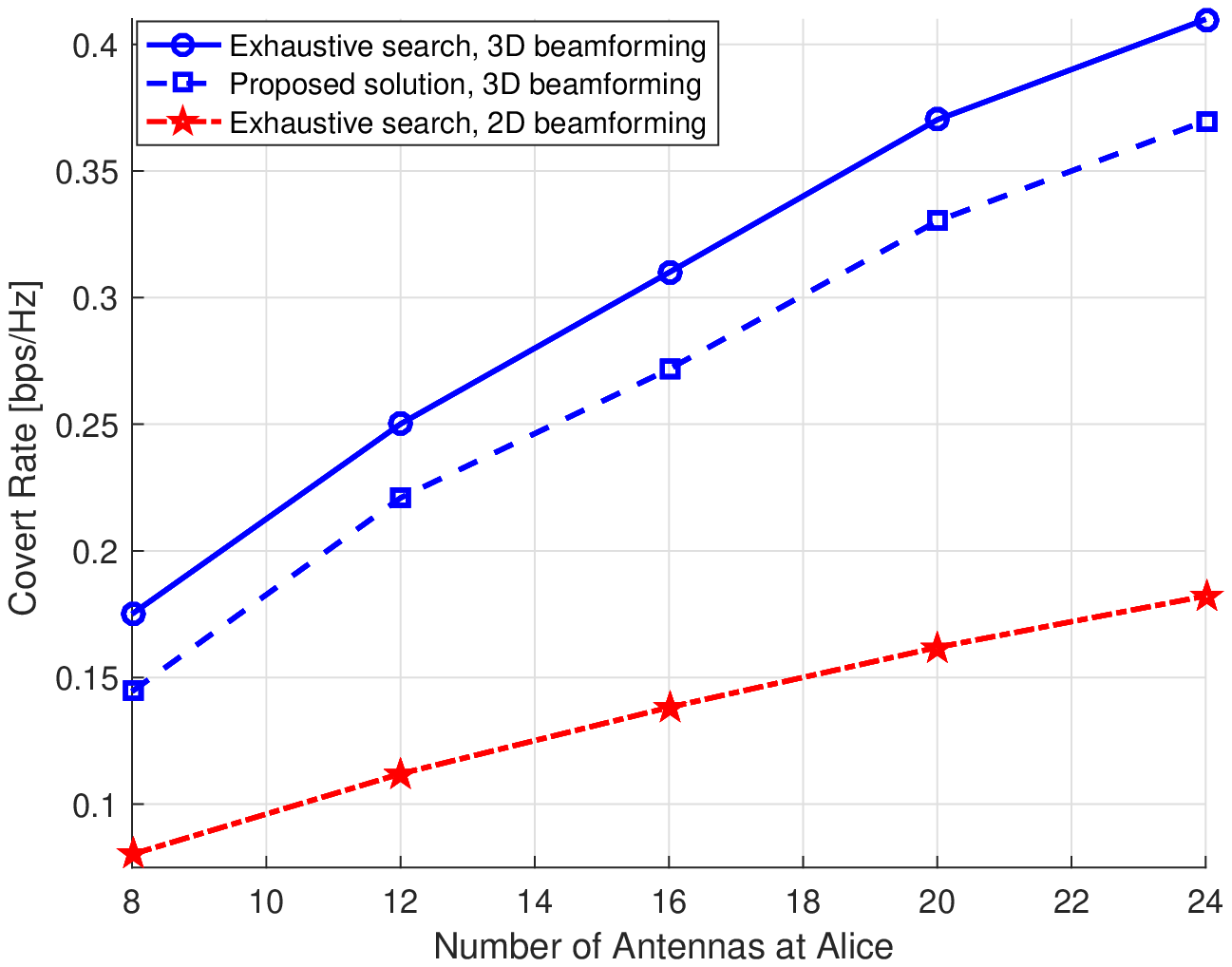}
			\label{Na}}
		
		\subfigure[h][Covert rate versus location of Willie, $N_a=8$.]{
			\includegraphics[width=2.4 in,height=1.8in]{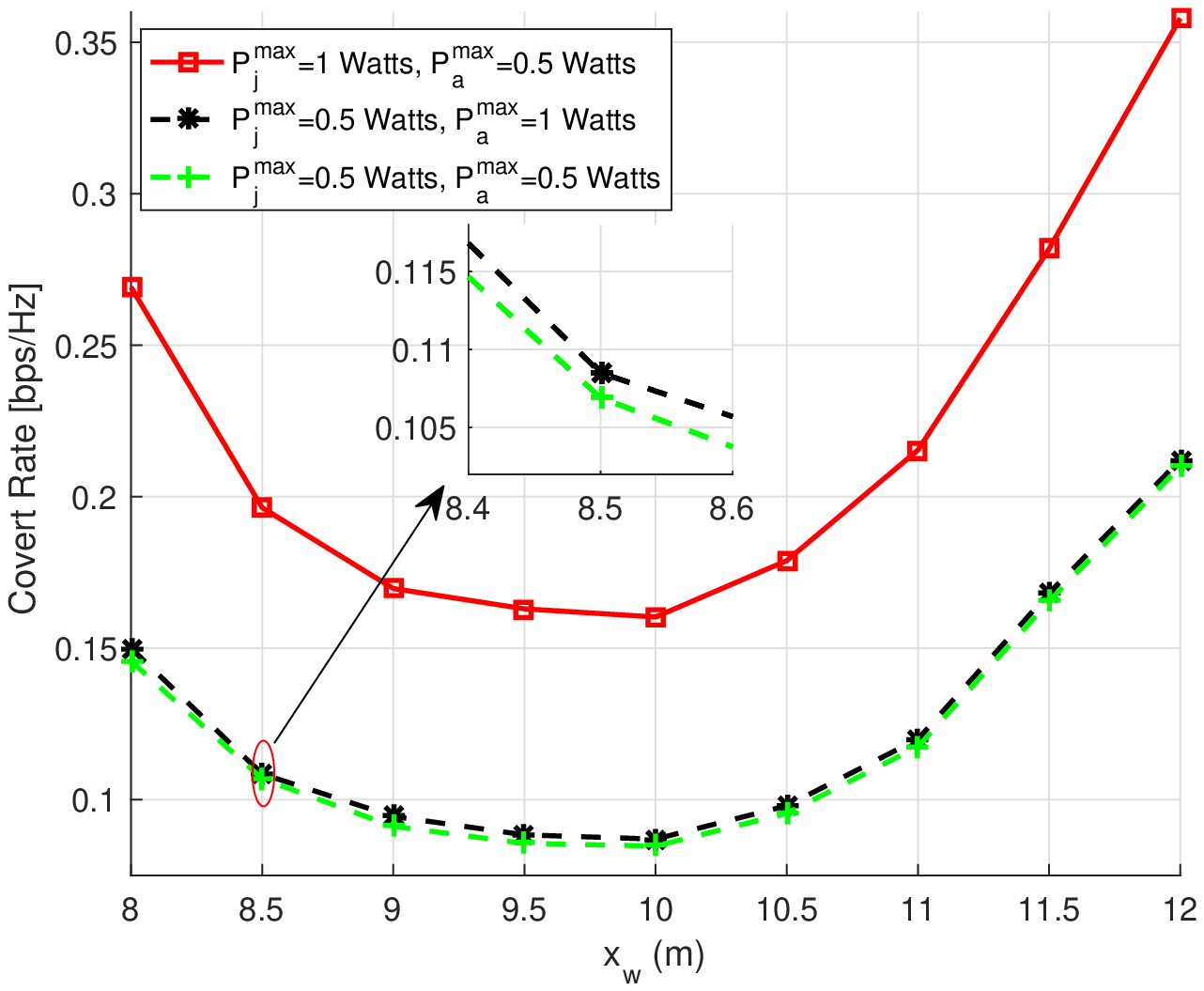}
			\label{xw}}

		\subfigure[h][Convergence of proposed algorithm \newline for $\tau=10^{-6}$.]{
			\includegraphics[width=2.4 in,height=1.8in]{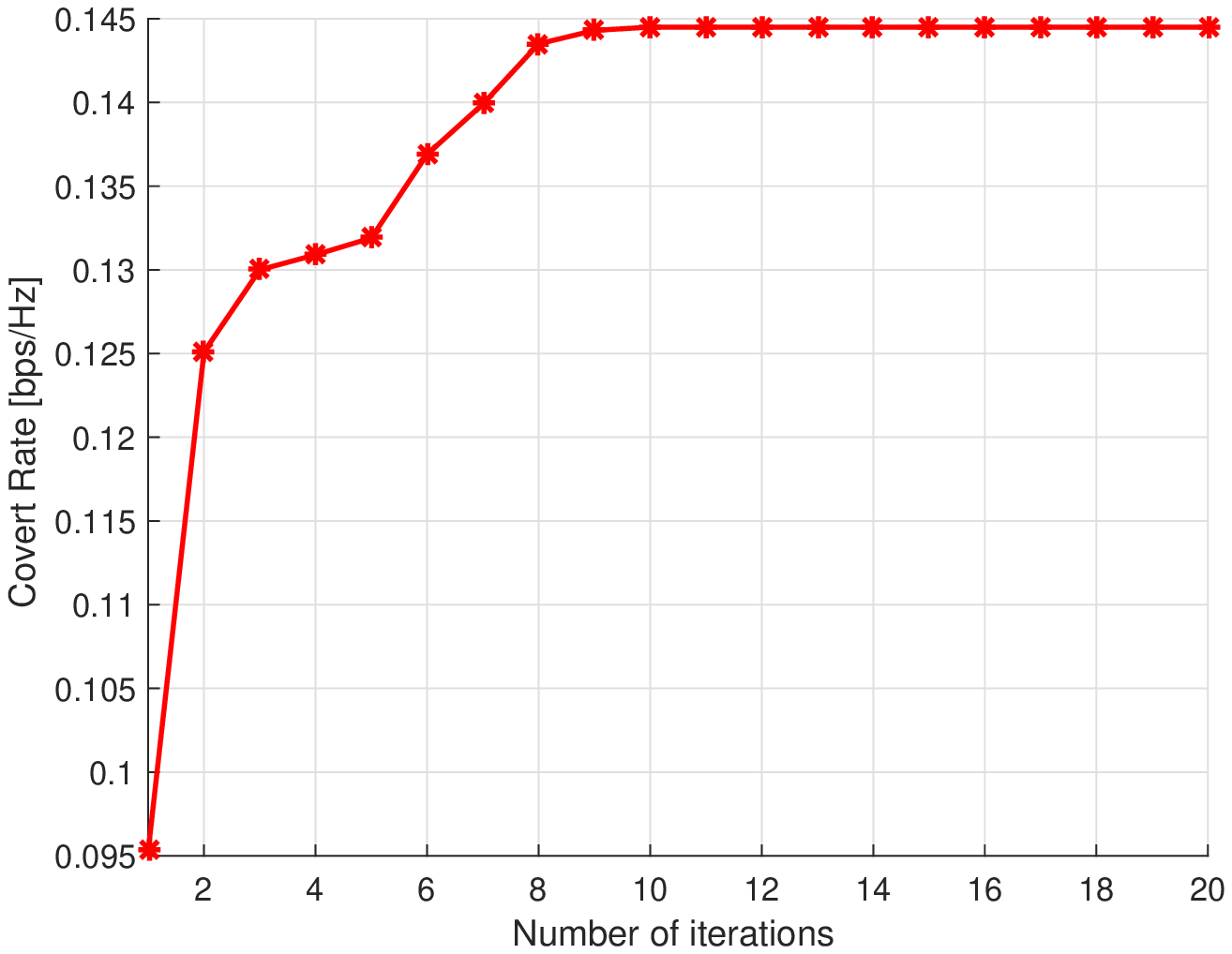}
			\label{Itr}}
		\end{array}$
	\end{center}\vspace{-.75cm}
	\caption{Evaluation of proposed network and proposed solution of optimization problem.}
	\label{SCW}
\end{figure*}

Fig. \ref{Na} shows the impact of the number of antennas at Alice on the covert rate when employing a 3D beamforming technique. This figure compares 2D and 3D beamforming in the proposed network. As seen, 3D beamforming compared to 2D beamforming increases covert rate about 50\%. Furthermore, this figure shows the optimality gap between the proposed solution and the optimal solution which is obtained  by the exhaustive search method in the 3D beamforming technique. As seen, this gap is about 12.4\%.

Fig. \ref{xw} evaluates the impact of the location of Willie on the covert rate when 3D beamforming technique is employed. It is worth noting that the worst case happens when Willie's location is near to that Bob. When Willi is near Bob, he can take advantage of the 3D beam from Alice. The covert rate is also plotted for different values of the maximum allowable transmit power of Alice and the jammer. As seen in this figure, increasing of the maximum allowable transmit power of the jammer has more effect on the covert rate compared to the maximum allowable transmit power of Alice. The reason is that by increasing  $P_j^{\max }$, Alice is able to greatly increase her transmit power while remaining covert. However, by increasing $P_a^{\max }$,  Alice can not  greatly increase the transmit power because in this case the covert communication cannot be guaranteed. 

Fig. \ref{Itr}  depicts convergence of the proposed Algorithm \ref{Alg_M1}. This figure is plotted for $N_a=8$, $ N_j=8$, and $P_a^{\max }=P_j^{\max }=0.5$ Watts. The proposed algorithm  converges in about $10$ iterations.

\section{Conclusion}\label{Conclusion}
Having multiple antennas at system nodes to steer the transmitted communication and jamming signals can obviously be of significant benefit to covert communications.  In this paper, we have considered how to optimize such a system to find the steering vectors and the resulting performance.  The pertinent optimization problem is non-convex, and thus we have proposed a sub-optimal iterative search to avoid the complexity of the exhaustive search.  Numerical results demonstrate that the performance of our sub-optimal algorithm is close to that obtained by exhaustive search.  The covert rate of the system employing 3D beamforming at transmitter Alice is roughly $50\%$ higher than that of a system where Alice uses 2D beamforming.  As expected, we demonstrate that system performance degrades due to a loss in beamforming gain if the adversary is close to the intended receiver Bob.  Finally, we observe that relaxing an average power constraint on Alice is not effective in improving the covert rate, as it is the covertness constraint that is active in our simulations.  Rather, increasing the jammer power, which allows Alice to transmit more power while remaining covert, is much more effective in improving performance.

\end{document}